\documentclass[prb,preprint]{revtex4}
\pdfoutput=1
\usepackage{graphicx}
\begin{document}


\title{Incommensurability of stripes in $\mathbf{La_{2-x}Sr_xNiO_{4+y}}$
\medskip }

\date{September 12, 2017} \bigskip

\author{Manfred Bucher \\}
\affiliation{\text{\textnormal{Physics Department, California State University,}} \textnormal{Fresno,}
\textnormal{Fresno, California 93740-8031} \\}

\begin{abstract}
An analytic expression for the incommensurability of static stripes in $La_{2-x}Sr_{x}NiO_{4+y}$ is given, depending on the hole density $n_h = x + 2y$.
Apart from geometry factors the formula is the same as for stripes in the related cuprates
$La_{2-x}Ae_{x}CuO_4$ ($Ae = Sr, Ba$). Agreement with experimental data from neutron and X-ray diffraction is good.
The stability of stripes is interpreted in terms of the separation of hole charges residing at every $\nu^{th}$ node of the associated magnetization waves.

\end{abstract}

\maketitle

Interest in magnetic density waves (MDWs) and charge-density waves (CDWs)
\linebreak ---collectively called ``stripes''---in strontium-doped and/or oxygen-enriched lanthanum nickelate, $La_{2-x}Sr_{x}NiO_{4+y}$ (LSNO), arose shortly after the discovery of high-$T_c$ superconduc- tivity in $La_{2-x}Ae_{x}CuO_4$ (LACO, $Ae = Sr, Ba$). The nickelate and the cuprates have the same crystal structure and they show stripes when hole-doped. Given these commonalities, the main motivation for the study of stripes in the nickelate has been to learn more about stripes in the cuprates, providing possible clues to the mechanism of the latter's high-$T_c$ superconductivity.\cite{1} A key quantity in that pursuit is the \emph{incommensurability} of the stripes (a wavenumber). Experimentally the incommensurability  $\delta(n_h, T)$, in its dependence on hole doping and temperature, is determined indirectly from the position of satellite peaks of neutron or hard X-ray diffraction,\cite{2,3,4,5,6,7,8,9,10,11} and directly with resonant soft X-ray scattering,\cite{12,13,14,15} as well as with electron microscopy.\cite{16} Despite a large amount of research over two decades no analytic expression for $\delta(n_h, T)$ has been put forward---other than the observation that in a certain doping range $\delta \approx n_h$. 
In this note a formula is presented for the doping dependence of the incommensurability of \emph{static} stripes, $\delta(n_h)$, in the nickelate LSNO. Apart from geometry factors the formula is the same as for stripes in the related cuprates LACO, extending its validity.\cite{17}

Hole doping of the parent compound $La_2NiO_4$ can be achieved through substitution of ionized lanthanum atoms, $La \rightarrow La^{3+} + 3e^-$, by ionized strontium atoms, $Sr \rightarrow Sr^{2+} + 2e^-$, in the $LaO$ layers  of the crystals. This causes electron deficiency (hole doping) of concentration $n_h = x$.
Each missing electron at the dopant site is replaced by an electron from an $O^{2-}$ ion, leaving an $O^-$ ion  behind. It is generally assumed that those $O^-$ ions, establishing the electronic ``holes,'' reside in the $NiO_2$ planes sandwiched by the $LaO$ layers.
Hole doping of $La_2NiO_4$ can also be achieved by incorporation of excess oxygen of concentration $y$ at interstitial sites. To attain (energetically favorable) closed-shell configuration, each interstitial oxygen atom takes two electrons from neighboring atoms, $O + 2e^- \rightarrow O^{2-}$, creating two holes, $n_h = 2y$. Accounting for both mechanisms, the hole density is given in general by
\begin{equation}
n_h = x + 2y \, .
\end{equation}

The charge and magnetic stripes in $La_{2-x}Sr_xNiO_{4+y}$ run diagonally with respect to the $Ni$-$O$ bonds. It has proved convenient to characterize the stripe wavevectors with a unit cell of double volume (space group $F4/mmm)$. This indexes the corresponding wavevectors as 
\begin{equation}
\mathbf{q_c} = (2\delta, 0, 1) \, , \,\,\,\,\,\,\,\,\,\,\,\,\,\,\,\,\,\,
\mathbf{q_m} = (1 \pm \delta, 0, 0) \, ,
\end{equation}
in reciprocal lattice units (r.l.u.).

Previously a formula for the incommensurability of MDWs and CDWs in the cuprates LACO was derived through partition of the $CuO_2$ planes by pairs of itinerant doped holes.\cite{18} 
Using the same assumptions for doped holes in the $NiO_2$ planes but accounting for the different indexing of stripes in the nickelate the incommensurability of stripes in LSNO is given as
\begin{equation}
\delta(n_h) = \frac{\sqrt{2}}{2} \sqrt{n_h - n_h^0} \,\, ,
\end{equation}
with $n_h = x + 2y$ from Eq. (1). The off-set by $n_h^0$ under the radical accounts for the density of holes at the emergence of stripes. Good overall agreement of the formula with experimental data (see Fig. 1) is achieved with $n_h^0 = 1/9 \simeq 0.111$. An interpretation of $n_h^0$ will be given below.

The data in Fig. 1 cluster about the $\delta(n_h)$ curve, except for the doping range 
\linebreak $0.20 \leq n_h \leq 0.25$ where they form a salient at $\delta \approx 0.27$. In the middle doping range, 
\linebreak $0.25 < n_h < 0.35$, the observed incommensurability approximately equals the hole doping, $\delta(n_h) \simeq n_h$---a relationship that was noticed in early research.\cite{2} However, for higher doping, $0.35 < n_h \leq 0.50$, there is a clear deviation of  $\delta(n_h)$ from linearity.

Also noticed in early research was that stripes are particularly stable when their incommensurability is close to commensurate values.\cite{19} The outstanding case is $\delta(\frac{1}{3}) = \frac{1}{3}$,
correctly accounted for by Eq. (3): 
$\delta(\frac{1}{3}) = \frac{\sqrt{2}}{2} \sqrt{\frac{1}{3} - \frac{1}{9}} =
\frac{\sqrt{2}}{2} \sqrt{\frac{2}{9}} = \frac{1}{3}$.
Graphically, the case is represented by the upper intersection of the $\delta(n_h)$ curve and diagonal in Fig. 1. 
It is noteworthy that experimental data are lacking by the lower intersection at $(n_h, \delta) = (\frac{1}{6},\frac{1}{6})$. This is, presumedly, not for lack of trying (to measure) but for lack of signal due to stripe instability. 
Only marginal stability of stripes exists in the doping range $0.20 \leq n_h \leq 0.25$ (mottled circles in Fig. 1),
as experimentally noticed by signals (peak heights) up to two orders weaker than for $\delta = 1/3$.\cite{15}

For an assessment of stripe stability we consider the spacing of the hole charges of CDWs residing at nodes of the associated MDWs. This is motivated by the stripes in the cuprates LACO where, according to $\delta_c = 2 \delta_m$, the doped charges always reside at \emph{every} node of the magnetization wave.\cite{18}
For simplicity we take into consideration only the projection of the density waves onto the $NiO_2$ plane, with wavelength $\lambda_c =1/(2\delta)$ for CDWs $\tilde{c}(\delta)$ 
and $\lambda_m^{\pm} =1/(1 \pm \delta)$ for MDWs 
$\tilde{m}^{\pm}(\delta)$ due to Eq. (2).
Each MDW, alternating sign according to the direction of magnetization, has two nodes per wavelength. The hole charge density of the associated CDW has one peak per wavelength, residing at magnetization  nodes, a number $\nu^{\pm} = \lambda_c/(\frac{1}{2} \lambda_m^{\pm})$ nodes apart,
\begin{equation}
 \nu^{\pm}(\delta) = \frac{1 \pm \delta}{\delta} \, .
\end{equation}

\includegraphics[width=6in]{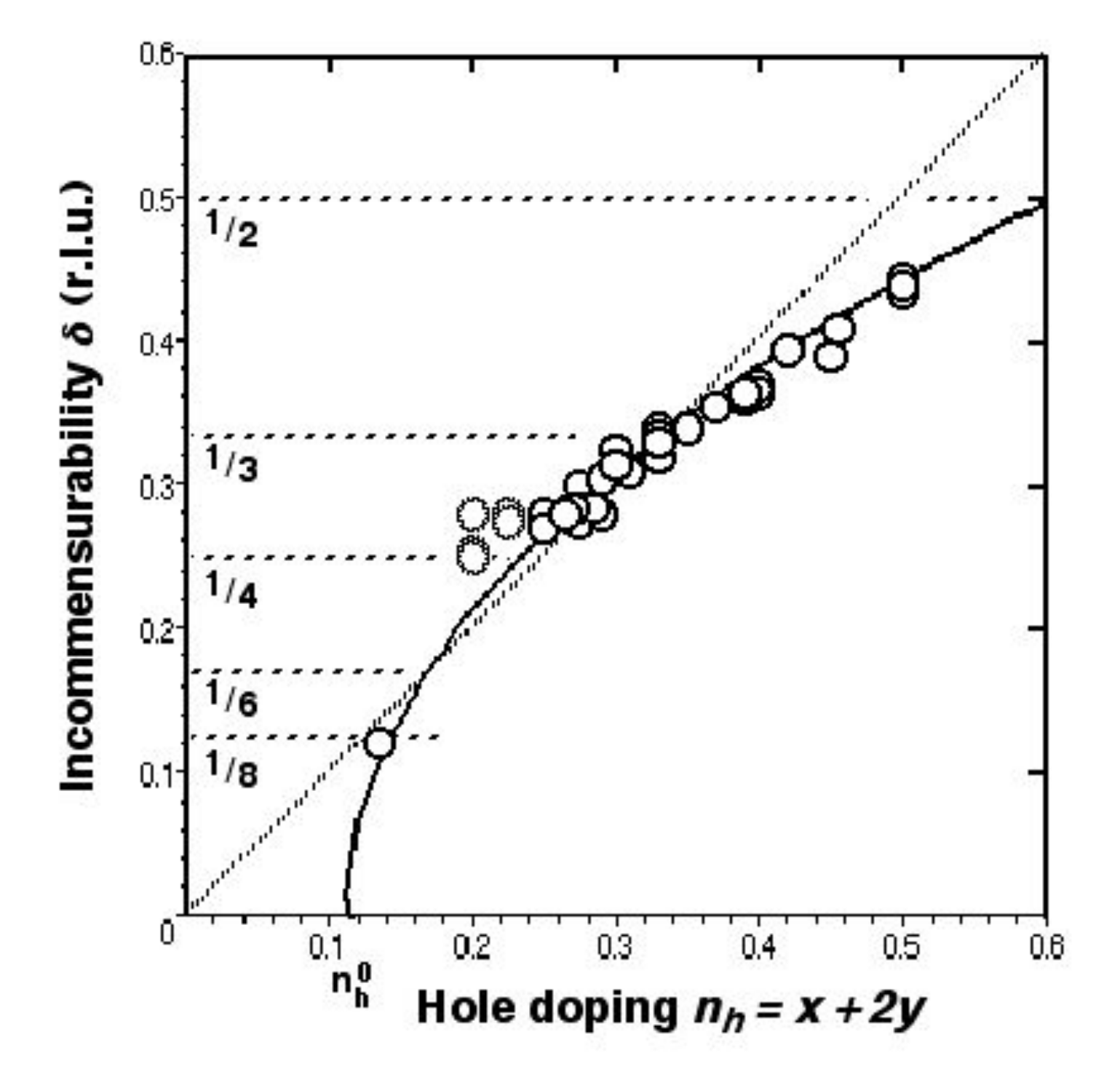}

\footnotesize 

\noindent FIG. 1. Incommensurability $\delta(n_h)$ of static stripes in $La_{2-x}Sr_{x}NiO_{4+y}$ due to hole doping by $n_h = x +2y$. Circles show data from neutron scattering  (Refs. 2-11), from X-ray diffraction (Refs. 12-15), and from electron microscopy (Ref. 16). The solid curve is a graph of Eq. (3). Dotted slanted line $\delta = n_h$, dashed horizontal lines at prominent commensurate values. Mottled circles in the doping range $0.20 \leq n_h < 0.25$ indicate very low stability of stripes.

\normalsize 

We now examine prominent cases. 
In the case of $\delta = \frac{1}{3}$ we obtain a separation of charged magnetization nodes by $\nu^-(\frac{1}{3}) = 2$ for the stripes formed by the 
$[\tilde{c}(\frac{1}{3}), \, \tilde{m}^-(\frac{1}{3})]$ pair. 
Accordingly, hole charges reside on every \emph{other} node of the magnetization wave. This gives these stripes in the nickelate LSNO a moderate stability. These are the stripes observed at and near $(n_h, \delta) = (\frac{1}{3},\frac{1}{3})$ (see Fig. 1).
Much lower is the stability of the other pair,
$[\tilde{c}(\frac{1}{3}), \, \tilde{m}^+(\frac{1}{3})]$, where $\nu^+(\frac{1}{3}) = 4$ manifests hole charges on every \emph{forth} node of the magnetization wave.

In the case $\delta = \frac{1}{4}$ the stripes formed by the $[\tilde{c}(\frac{1}{4}), \, \tilde{m}^-(\frac{1}{4})]$ pair have a separation of charged magnetization waves by  $\nu^-(\frac{1}{4}) = 3$. The stability of these stripes is marginal as observed for data in the interval $0.20 \leq n_h \leq 0.25$ with $\delta \approx \frac{1}{4}$ (mottled circles in Fig. 1). 
For $\delta = \frac{1}{6}$, corresponding to the lower intersection of the $\delta(\nu)$ curve and diagonal in Fig. 1, $\nu^-(\frac{1}{6}) = 5$ and $\nu^+(\frac{1}{6}) = 7$. The wide separation of hole charges at every fifth or seventh magnetization node, respectively, renders these stripes unstable. This accounts for the lack of data at and near $(n_h, \delta) = (\frac{1}{6},\frac{1}{6})$.

Turning to the hypothetical case $\delta = \frac{1}{2}$, Eq. (4) gives $\nu^-(\frac{1}{2}) = 1$, corresponding to hole charges at \emph{every} node of the magnetization wave, as in the cuprates LACO and resulting in the highest stripe stability. The incommensurability of the data in the $0.40 < n_h \leq 0.50$ doping range is $\delta(n_h) < \frac{1}{2}$. Nevertheless, the proximity to $\delta(n_h) = \frac{1}{2}$ seems to lend some stability. For the data point $(n_h, \delta) = (0.50, 0.44)$ Eq. (4) gives a separation of charged holes at $\nu^-(0.44) = 1.27$ nodes of the associated magnetization wave. Curiously, the case of stripes with the highest stability---node separation of $\nu^-(\frac{1}{2}) = 1$---at a hole doping of

\begin{table}[h!]
\begin{tabular}{ |p{8cm}|p{4cm}|  }
 \hline  \hline
Separation of charged magnetization nodes & Stability of stripes \\
\hline
$\nu^- = 1$ \,\,\,\,\,\,\,\,\,\,\,\,\,\,\,\,\, at $\delta = 1/2$ & high \\
 \hline 
 $\nu^- = 2$ \,\,\,\,\,\,\,\,\,\,\,\,\,\,\,\,\, at $\delta =  1/3$ & moderate \\
 \hline
 $\nu^- = 3$ \,\,\,\,\,\,\,\,\,\,\,\,\,\,\,\,\, at $\delta = 1/4$ & marginal \\
 \hline
 $\nu^- \geq 4$ \,\,\,\,\,\,\,\,\,\,\,\,\,\,\,\,\,\, at $\delta = 1/6$ & unstable \\
 \hline
 $\nu^{\pm} \approx 8 \gg 1 $ \,\,\,\,\,\, at $\delta = 1/8$ & weakly stabilized by \\
  & 1/8 commensuration \\
 \hline   \hline
\end{tabular}
\caption{Qualitative assessment of stripe stability in the nickelate LSNO depending on hole charges residing at every $\nu^{th}$ node of the associated magnetization wave, with $\nu^{\pm}$ from Eq. (4).}
\label{table:1}
\end{table}

\noindent  $n_h \simeq 0.60$ has not been observed in the nickelate LSNO (see Fig. 1).

The last case to be discussed is for sufficiently small incommensurability, $\delta = 1/N$, when $N \gg 1$. Then Eq. (4) becomes
$\nu^{\pm}(\delta) = (1/\delta) \pm 1 \approx N$. The distinction between the $\tilde{m}^-(\delta)$ and $\tilde{m}^+(\delta)$ MDWs then practically ceases. Although the hole charges reside far apart near every $N^{th}$ magnetization node, the stability of the MDW is assisted by approximate $1/N$ commensuration with the lattice. This may be the case for the lone data point at $n_h = 0.135$ with $\delta = 0.12 \simeq 0.125 = 1/8$ (see Fig. 1).
Table I gives a qualitative assessment of stripe stability by separation of charged magnetization nodes.

For an interpretation of the off-set $n_h^0 = 1/9$ in the radicand of Eq. (3) we draw on analogies with the cuprates LACO. There the off-set $x_0^N = x_{10} \equiv 2/10^2 = 0.02$ is the concentration of $Ae = Sr$ or $Ba$ doping where the N\'{e}el temperature vanishes, $T_N(x_0^N)=0$. This causes (at $T=0$) collapse of the insulating, long-range antiferromagetic (3D-AFM) phase with subsequent emergence of a metallic, so-called ``pseudogap'' phase. The $x_0^N$ presence of holes at $O^{-}$ ions in the $CuO_2$ planes keeps the 3D-AFM phase suppressed. The remaining holes, of concentration $x - x_0^N$, give rise to the combined CDWs/MDWs (stripes) while the unaffected $O^{2-}$ ions, together with the $Cu^{2+}$ spin magnetic moments, maintain 2D-AFM, sometimes called ``spin glass.''

The 3D-AFM phase in the nickelate LSNO is considerably more stable than in the cuprates LACO due to spin magnetism with $S=1$ of the $Ni^{2+}$ $3d^8$ ions compared to $S=1/2$ of the $Cu^{2+}$ $3d^9$ ions. Considering only the magnetic spin-spin interaction of energy $E_{ss}^m \propto S^2$ gives a crude estimate for the increased concentration of doped holes in the nickelate to suppress 3D-AMF, $\widetilde{n_h^0} \approx 4 x_0^N = 0.08$. 
Whereas the N\'{e}el temperature in the alkaline-earth doped cuprates, $La_{2-x}Ae_xCuO_4$, decreases linearly with increasing hole concentration, $T_N(x) =- [T_N(0)/x_0^N]x + T_N(0)$, a complicated $T_N(y)$ profile is observed in oxygen-enriched $La_2NiO_{4+y}$, with a steep ``leg'' of linear descent, 
from $T_N = 333$ K at $y = 0$ to  $T_N = 160$ K at $y = 0.03$, but a long ``foot'' with $T_N \approx 80$ K from $y = 0.06$ to $y \approx 0.12$.\cite{20}
It is conceivable that the ``foot anomaly'' of $T_N(y)$ in $La_2NiO_{4+y}$ is a result of the excess oxygen, but that  $T_N(x)$ in strontiuum-doped $La_{2-x}Sr_xNiO_4$ would decrease linearly all the way as in the cuprates LACO. 
But it is also possible that the anomalous antiferromagnetic pattern is caused by the transition of low-temperature crystal phases, characterized by different tilting of $NiO_6$ octahedra, near $T = 70$ K.\cite{20} 
It is noteworthy, though, that a linear extrapolation of the $La_2NiO_{4+y}$ high-temperature $T_N(y)$ data intersects the $T=0$ axis at $\grave{y} \simeq 0.055$. This corresponds, by Eq. (1), to an intercept
$\grave{n}_h \simeq 0.11$, very close to the off-set $n_h^0 = 1/9 \simeq 0.111$ and somewhat larger than the estimate $\widetilde{n_h^0} \approx  0.08$.

One more lesson that can be learned from the cuprates LACO is that special doping concentrations are observed with an inverse-square dependence on integer numbers, $x_n = 2/n^2$.\cite{16} Prominent cases are (i) the off-set at the N\'{e}el point, $x_0^N = x_{10} = 0.02$ as mentioned, (ii) the change of stripe orientation from diagonal to parallel at $x_6 = 0.056$ as well as the onset of superconductivity at $x_0^{SC} = x_6$, and (iii) the dip in the $T_c(x)$ profile at $x_4 = 1/8 = 0.125$. 
(The reason for the inverse-square dependence of $x_n$ is not clear.)
The off-set for the nickelate, $n_h^0 = 1/9 = 1/3^2$, may similarly be a special case.

\bigskip \bigskip 

\centerline{ \textbf{ACKNOWLEDGMENTS}}

\noindent I thank Duane Siemens for valuable discussions and Preston Jones for help with LaTeX.

\end{document}